# North–South Asymmetries in Earth's Magnetic Field

## Effects on High-Latitude Geospace

K.M. Laundal[1,2]  · I. Cnossen[3] · S.E. Milan[4,1] ·
S.E. Haaland[5,1] · J. Coxon[6] · N.M. Pedatella[7] ·
M. Förster[8] · J.P. Reistad[1]

**Abstract** The solar-wind magnetosphere interaction primarily occurs at altitudes where the dipole component of Earth's magnetic field is dominating. The disturbances that are created in this interaction propagate along magnetic field lines and interact with the ionosphere–thermosphere system. At ionospheric altitudes, the Earth's field deviates significantly from a dipole. North–South asymmetries in the magnetic field imply that the magnetosphere–ionosphere–thermosphere (M–I–T) coupling is different in the two hemispheres. In this paper we review the primary differences in the magnetic field at polar latitudes, and the consequences that these have for the M–I–T coupling. We focus on two interhemispheric differences which are thought to have the strongest effects: 1) A difference in the offset between magnetic and geographic poles in the Northern and Southern Hemispheres, and 2) differences in the magnetic field strength at magnetically conjugate regions. These asym-

KML, SEM, SH, and JPR were supported by the Research Council of Norway/CoE under contract 223252/F50. IC was supported by a fellowship of the Natural Environment Research Council, grant number NE/J018058/1. NP was supported by the U.S. National Science Foundation AGS-1522830. JCC was funded by Natural Environment Research Council (NERC) grant NE/L007177/1. We acknowledge the International Space Science Institute for support for our international team on "Magnetosphere–ionosphere–thermosphere coupling: differences and similarities between the two hemispheres."

✉ K.M. Laundal
  karl.laundal@ift.uib.no

[1] Birkeland Centre for Space Science, University of Bergen, Bergen, Norway

[2] Teknova AS, Kristiansand, Norway

[3] British Antarctic Survey, Cambridge, UK

[4] University of Leicester, Leicester, UK

[5] Max-Planck Institute for Solar Systems Research, Göttingen, Germany

[6] Department of Physics and Astronomy, University of Southampton, Highfield, Southampton, SO17 1BJ, UK

[7] COSMIC Program Office, University Corporation for Atmospheric Research, Boulder, CO, USA

[8] German Research Centre for Geosciences, Helmholtz Centre Potsdam, Potsdam, Germany

metries lead to differences in plasma convection, neutral winds, total electron content, ion outflow, ionospheric currents and auroral precipitation.

## 1 Introduction

There are significant differences between the Earth's magnetic field in the Northern and Southern polar regions, even when seen in a magnetic field-aligned coordinate system. The magnetic flux density at magnetically conjugate points can differ by up to a factor of 2 at 50° magnetic latitude, and the absolute inclination angle by more than 10°. In addition, the magnetic apex pole is more than 8.5° farther from the geographic pole in the Southern Hemisphere (SH) compared to the Northern Hemisphere (NH), which means that the polar region in the South experiences a larger daily variation in sunlight as the Earth rotates. The longitudinal variation in magnetic flux density and field inclination is also much larger in the SH. These asymmetries between the hemispheres lead to differences in ionospheric plasma convection, auroral intensity, thermospheric wind, total electron content, and magnetic field perturbations and associated currents. In this paper we review the differences in the magnetic field at polar latitudes in the two hemispheres, and describe in detail how they may lead to differences in geospace activity.

The degree of inter-hemispheric symmetry depends on the reference frame which is used. A number of magnetic coordinate systems exist, taking into account the structure of Earth's magnetic field at different levels of detail. The most advanced magnetic coordinate systems, the corrected geomagnetic (CGM) coordinates (e.g., Baker and Wing 1989) and apex coordinates (Richmond 1995b), are based on tracing along magnetic field lines in the International Geomagnetic Reference Field (IGRF) model (Thébault et al. 2015) at full resolution. They are designed such that points that belong to the same field line are at the same coordinate, with a change of sign in latitude between hemispheres. A map of Modified Magnetic Apex coordinates is shown in Fig. 1. Note that the coordinate grid is nonorthogonal. This is an effect of the non-dipole terms of the IGRF; if they were zero and the Earth spherical, apex coordinates would be equal to the simpler centered dipole coordinate system. We use apex or CGM coordinates, which are similar at high latitudes, throughout this paper, since the field-aligned property implies that disturbances created by solar wind-magnetosphere interaction or magnetotail processes most often appear at the same magnetic coordinate in the two hemispheres, since the coupling between the ionosphere and magnetosphere is largely field-aligned.

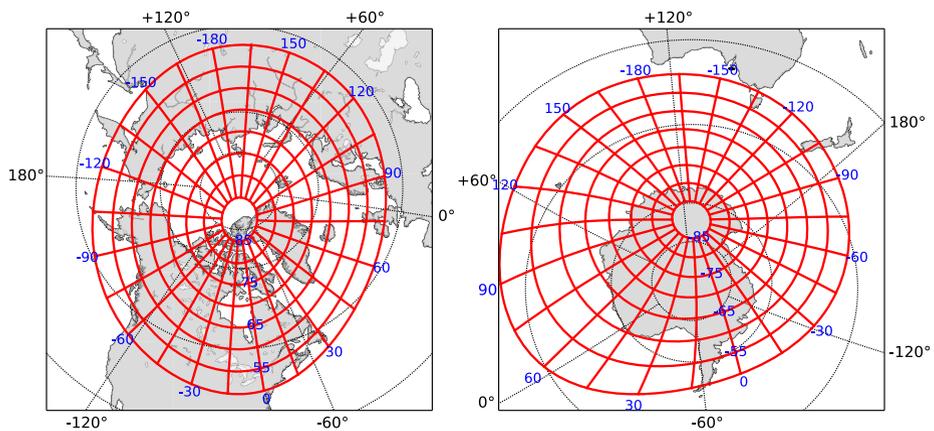

**Fig. 1** Modified apex coordinates (Richmond 1995b; Emmert et al. 2010), with reference height equal to 0. Adapted from Laundal and Gjerloev (2014)

The IGRF can be seen as a ground state of the magnetic field in the magnetosphere, which in reality is never reached at high altitudes: the solar wind-magnetosphere interaction compresses the magnetosphere on the day-side and creates the magnetotail on the night-side. Ono (1987) showed that this effect, during geomagnetic quiet times, creates a daily variation in the location of magnetically conjugate points at high latitudes. The variation at the Syowa station (at $\approx -66°$ CGM latitude) was approximately 100 km during solstices, and much less at equinox. In addition, the interaction of the magnetosphere with the solar wind and the ionosphere–thermosphere system is often asymmetrical between hemispheres, twisting the magnetosphere such that magnetically conjugate phenomena appear shifted in longitude and/or latitude. Such shifts, which have been observed to reach $\approx 2$ hours of magnetic local time (Østgaard et al. 2011), have been extensively studied, and we will not go into details in this paper. When we talk about asymmetries in the magnetic field at conjugate points, we refer to their position according to the IGRF.

The two features of the asymmetric magnetic field which are probably most important for geospace phenomena are the field strength asymmetries at conjugate points and the differences in offset between the magnetic and geographic grids. The differences in offset between magnetic and geographic coordinates imply that the interaction between magnetically and geographically organized phenomena will be different in the two hemispheres. The latter includes the exposure to sunlight, which largely determines the ionospheric conductivity on the day-side, and consequently also the strength of thermosphere–ionosphere coupling. A given point in the SH will in general experience larger variations in sunlight throughout a day compared to its conjugate point in the NH.

Differences in field strength mean that the mirror height of trapped charged particles will be different. Where the field is weak, the mirror height is lower, suggesting that more particles will interact with the atmosphere there and create ionization and auroral emissions. However, the area over which the precipitating particles are distributed will be larger at regions with lower flux density, and thus the intensity will be lower. Whether or not the mirror height effect and the differences in area balance depends on the pitch angle distribution of the particles (Stenbaek-Nielsen et al. 1973). This will be treated in more detail in Sect. 8. Differences in magnetic flux density also affect the ionospheric conductance, which is inversely related to the magnetic field strength (Richmond 1995a; Cnossen et al. 2011, 2012a). This may have important effects on ionospheric currents and associated magnetic field disturbances, as well as the plasma flow (Cnossen et al. 2011, 2012a). The response of the ionosphere to magnetospheric driving depends on the Pedersen conductance (e.g. Scholer 1970), or ionospheric mass (Tu et al. 2014), suggesting that the magnetosphere–ionosphere coupling may be different in the two hemispheres and at different longitudes. Modeling by Förster and Cnossen (2013) has indeed shown that the asymmetric features in the Earth's field introduces differences in plasma convection and thermospheric winds at high latitudes.

In Sect. 2 we present a detailed description of the asymmetric features in the magnetic field in the two hemispheres. The subsequent sections explore the effects of these asymmetries on plasma drift (Sect. 3), thermospheric wind (Sect. 4), total electron content (Sect. 5), ion outflow (Sect. 6), currents and magnetic field perturbations (Sect. 7), and the aurora (Sect. 8). Section 9 concludes the paper.

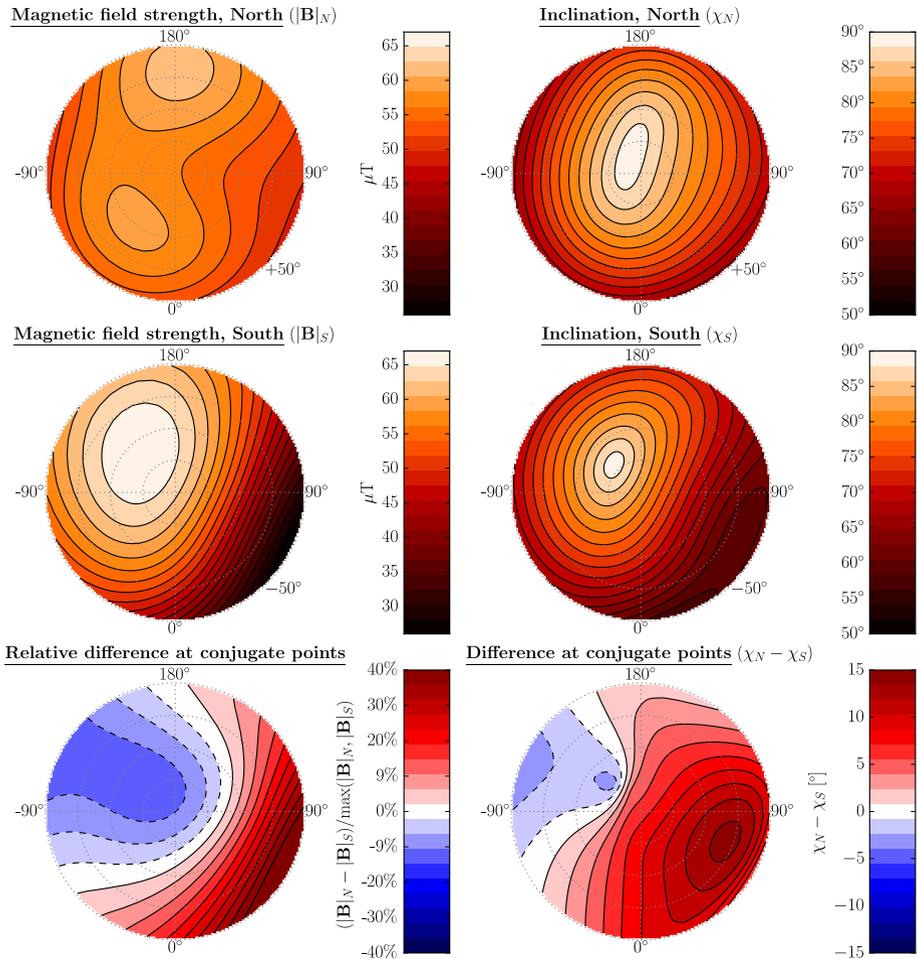

**Fig. 2** Magnetic field strength (*left column*) and absolute inclination (*right column*) in apex coordinates in NH (*top*), SH (*middle*) and the difference between the hemispheres (*bottom*). The inter-hemispheric difference in field strength is shown relative to strongest field among the two footpoints. IGRF-12 values for 2015 were used, at 1 Earth radius

## 2 North–South Magnetic Field Asymmetries at High Latitudes

### 2.1 Magnetic Field Strength Differences at Conjugate Points

Figure 2 shows the ground magnetic field strength (left column) and absolute inclination angle (right column) in the NH (top) and SH (middle), in the apex quasi-dipole (QD) coordinate system. The bottom row shows the inter-hemispheric difference in these quantities. The difference in magnetic field strength is quantified as the hemispheric difference divided by the flux density at the footpoint with the strongest field. Positive values signify stronger field values in the NH. The asymmetry in field inclination at conjugate points is quantified as the difference between the angles, positive where the field is closest to vertical in the NH.

We see that the flux density is more uniform in the NH than the SH. The field in the NH has two maxima, located in the Canadian and Siberian sectors (around $-30°$ and $180°$ magnetic longitude, respectively). In the SH the field has only one maximum, off the apex pole towards Australia (at $\approx -135°$ longitude), and decreases significantly towards the South Atlantic region. The difference at conjugate points at Atlantic longitudes is up to a factor of 2. In the polar cap region poleward of $\approx \pm 80°$, the field is stronger in the SH by approximately 7%. Equatorward of this, the field is strongest in the NH everywhere except for the quadrant between $-90°$ and $180°$ magnetic longitude.

The Hall and Pedersen conductivities depend on the magnetic field strength directly and via its effect on electron and ion gyro frequencies (Richmond 1995a). The height integrated dayside conductances were reported by Richmond (1995a) to scale with $B^{-1.3}$ (Hall) and $B^{-1.6}$ (Pedersen). Later modeling results, investigating the change on the coupled magnetosphere–ionosphere–thermosphere system associated with a changing dipole moment, have shown larger scaling factors: Cnossen et al. (2011) found scaling factors of approximately $B^{-1.7}$ (Hall) and $B^{-1.5}$ (Pedersen) on the dayside. They also found a variation with $B$ on the nightside, but significantly smaller. In a later study Cnossen et al. (2012a) found that the variation of the Pedersen conductance with magnetic field strength is stronger when the solar EUV flux is higher.

Using the comparatively moderate scaling parameters from Richmond (1995a), we find that a relative difference of $\pm 20\%$ in magnetic flux density amounts to a relative difference in Hall conductance of approximately $\mp 25\%$ (notice the change in sign) and Pedersen conductance of approximately $\mp 30\%$. Differences of this magnitude or larger occur up to $70°$ magnetic latitude in the $0°$–$90°$ longitude quadrant.

The inclination or dip angle of the magnetic field is also different in the two hemispheres. The hemispheric difference follows approximately the same pattern as for the magnetic field strength, with the field lines in the NH more vertical in the regions where the field is strongest. The asymmetry reaches a peak in the $0°$–$90°$ longitude sector, where the difference reaches more than $10°$ at latitudes just poleward of $\pm 65°$.

Figure 3 illustrates the longitudinal variation of the magnetic field in both hemispheres. The left part shows the relative difference between the strongest and weakest field values along circles of constant magnetic latitude (maximum divided by minimum), given on the $x$ axis. The dashed and dotted curves show the corresponding relative differences in Pedersen and Hall conductances, assuming that they scale as $B^{-1.6}$ and $B^{-1.3}$, respectively. We see that in the SH, the magnetic flux density varies by more than a factor of 2 at $55°$ latitude. The corresponding variation in daytime Pedersen conductance is approximately a factor of 3.5 and Hall conductance close to 3. In the NH, the magnetic field is much more uniform, the relative longitudinal variation in flux density at $>50°$ being approximately 1.25 at most. These inter-hemispheric differences, together with larger daily variation in solar illumination, are likely to produce larger diurnal variations in geomagnetic activity in the SH compared to the NH.

The right part of Fig. 3 shows the longitudinal variation in magnetic inclination angle. In this figure we show the absolute variation rather than relative variation. The difference in the inclination angle along a circle of latitude reaches $7°$ in the NH and $18°$ in the SH.

### 2.2 Differences in Pole Offsets

Figure 4 illustrates the variation in sunlight exposure on the magnetic grids in the two hemispheres. The upper part of the figure shows apex quasi-dipole circles of latitude at $\pm 60°$,

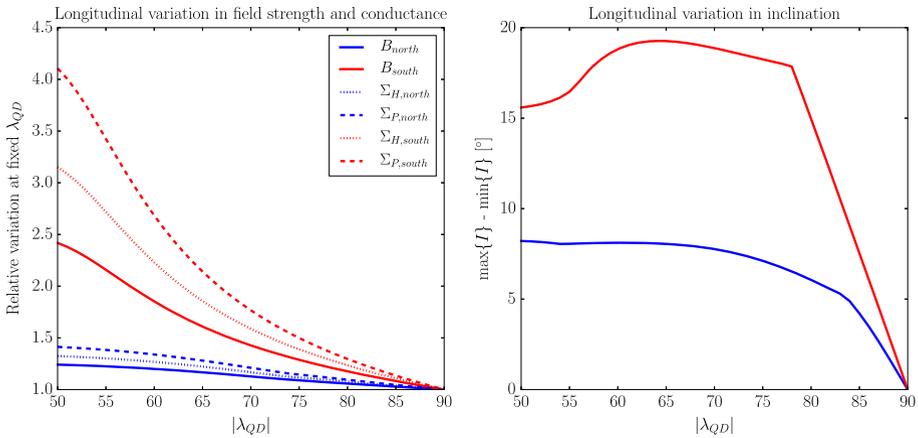

**Fig. 3** *Left*: Relative variation in $B$, $\Sigma_H$ and $\Sigma_P$ (assuming they scale as $B^{-1.3}$ and $B^{-1.6}$, respectively (Richmond 1995a)) in both hemispheres. The relative variation is quantified as the maximum value divided by the minimum value along a contour of constant magnetic latitude, given at the $x$ axis. *Right*: Longitudinal variation in magnetic inclination as a function of magnetic latitude. IGRF-12 values for 2015 were used

±70° and ±80° in both hemispheres projected on a geographic grid in the NH. In addition, magnetic meridians separated by 90° are shown, with the 0° meridians in bold. Blue color corresponds to the NH and red to the SH. The markers signify conjugate points at which magnetometer stations are located (to be discussed in Sect. 7). The offset between the magnetic and geographic poles is clearly seen. Due to the offset between geographic and magnetic poles, there will be certain universal times when one hemisphere (in magnetic coordinates) is more sunlit than the other. The panel in Fig. 4 shows this UT variation, quantified in terms of the fraction of the region poleward of ±60° which is sunlit. Positive hemispheric differences mean that the NH is more sunlit than the SH. This figure corresponds to equinox conditions, but the general UT variation will be similar in other seasons.

The lower plot illustrates how the exposure to sunlight varies throughout the year in the regions poleward of 60° magnetic latitude. The curves, blue for the NH and red for the SH, show the daily minimum and maximum fraction of the region poleward of 60° which is sunlit. Since the distance between these curves is larger in the South than in the North, the daily variation is always largest in the South. Notice that the polar circle (black dashes in the top left plot), which is tangent to the sunlight terminator at solstice, is equatorward of the −60° QD latitude contour. That means that at certain UTs, the region at < −60° will be entirely dark (sunlit) close to Southern winter (summer), so that the sunlit fraction envelope curve saturates at 0 (1).

At solstice, there is naturally a large difference in solar illumination between the summer and winter hemispheres. To eliminate North–South differences arising simply from this effect, it can be helpful to compare the two hemispheres in the same local season (e.g., winter or summer). Even then though, there are small differences in the amount of solar radiation received by the Northern and Southern Hemispheres. This is partly due to the different offsets between the geographic and magnetic poles, which result in differences in solar zenith angle and length of day, and partly due to the elliptical shape of the Earth's orbit around the Sun, which results in variations in Sun–Earth distance over the course of the year. The

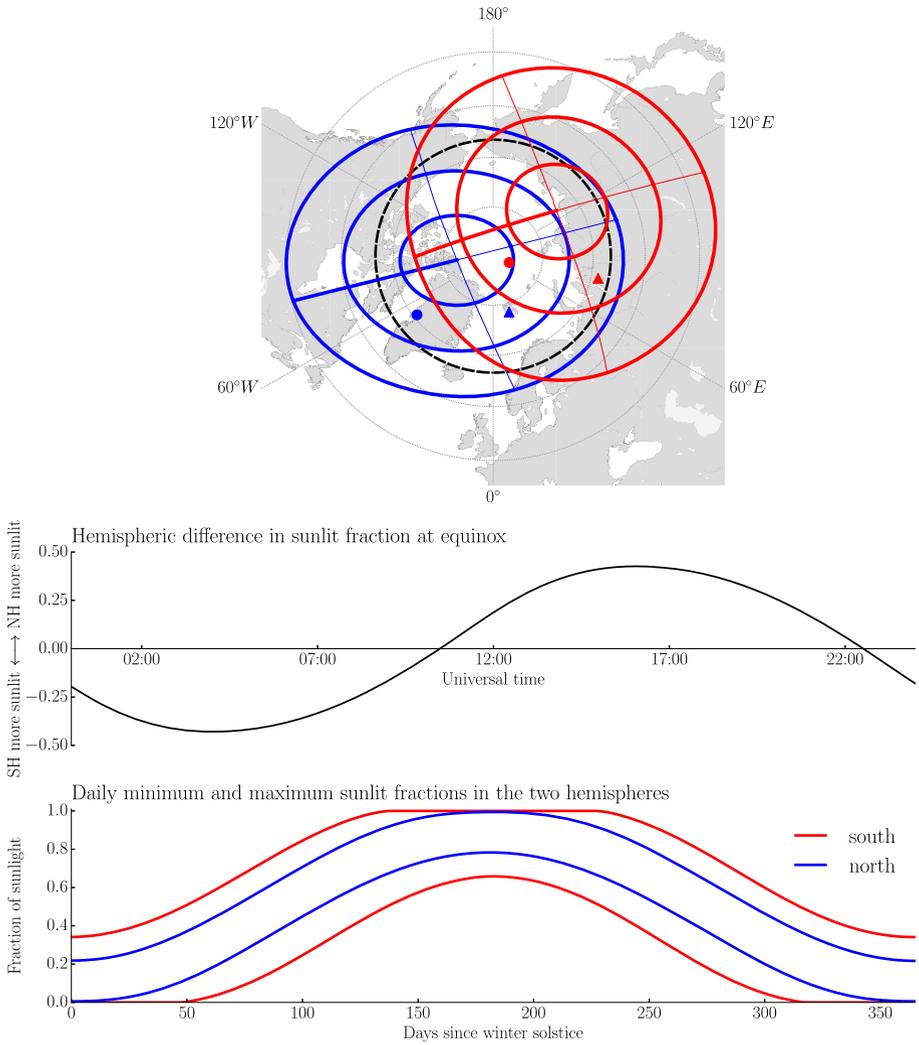

**Fig. 4** *Top*: Magnetic QD grids from both hemispheres (*red* is South, *blue* is North) shown in geographic coordinates, projected to the NH. The ±60°, ±70° and ±80° circles of latitude are shown. The 0° magnetic meridian is shown in *bold*. The *circles* and *triangles* mark conjugate magnetometers, discussed in Sect. 7. The polar circle at 66.6° is shown in *black dashes*. *Middle*: The hemispheric difference in the fraction of the region poleward of ±60° QD latitude which is sunlit. The *curve* represents equinox conditions. Positive values mean that the NH is more sunlit than the SH. *Bottom*: The minimum and maximum fractions of the region poleward of 60° which is sunlit as a function of days since the last local winter equinox

Earth is about $5 \cdot 10^6$ km closer to the Sun in early January (perihelion) than in early July (aphelion), causing a difference in illumination of about 6–7%.

Figure 5 shows the mean daily insolation at the apex magnetic poles in the Northern and Southern Hemisphere as a function of days since winter solstice, both with and without the effect of the variation in Sun–Earth distance. While the higher geographic latitude of the apex magnetic pole in the NH results in a larger solar zenith angle, the day is also longer in summer, so that the effect of the difference in offset between the magnetic and geographic poles is to result in greater insolation at the Northern apex pole. However, the effect of the variation in Sun–Earth distance is more important and reverses the asymmetry, so that on balance, the Southern apex pole receives more sunlight during most local seasons (except for a period of about a month in spring).

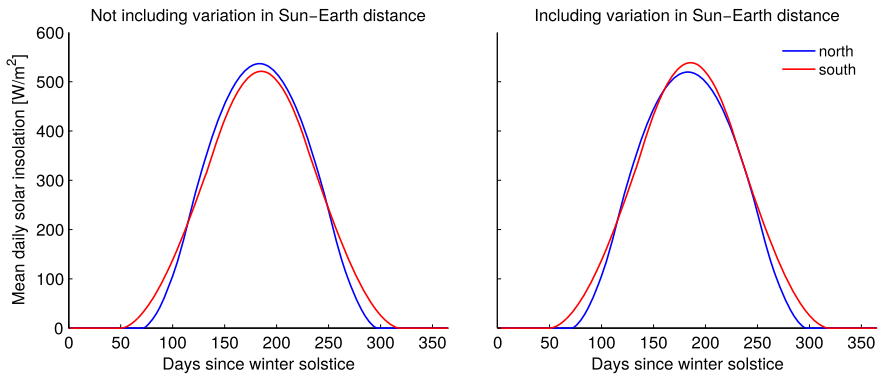

**Fig. 5** Mean daily solar insolation at the locations of the apex magnetic poles in the Northern (*blue*) and Southern (*red*) Hemisphere as a function of days since winter solstice, assuming a solar constant of 1361 W/m$^2$. *Left*: not including the effect of variation in Sun–Earth distance; *right*: including the effect of variation in Sun–Earth distance

## 3 Asymmetry Effects on Ionospheric Plasma Convection

When the interplanetary magnetic field (IMF), embedded in the solar wind plasma has a southward component, magnetic reconnection on the dayside magnetopause changes the field topology such that the closed terrestrial field lines become connected to the Sun's magnetic field, forming the polar caps. The polar caps are regions threaded by equal amounts of open magnetic flux in the two hemispheres. Being connected to the solar wind, the open field lines in the magnetosphere are transported anti-sunward, and folded into two so-called lobes in the magnetotail. In this process, solar wind kinetic energy is converted to magnetic energy which resides in the lobes until magnetic reconnection on the nightside creates new closed field lines. As these newly closed field lines relax from their highly stretched configuration, the magnetic energy is converted back to kinetic energy and a large-scale sunward plasma flow on closed field lines takes place. Eventually the field lines will end up on the dayside, where reconnection with the IMF starts a new cycle of plasma and magnetic flux circulation.

The footprint of this circulation, which is called the Dungey cycle (Dungey 1961), can be observed in the ionosphere as a two-cell flow pattern of ionospheric plasma. The ionospheric plasma flow is anti-sunward across the polar cap. The plasma then leaves the polar cap through the segment of its boundary that maps to the nightside reconnection region, before turning sunward on the dawn and dusk flanks, and eventually re-entering the polar cap in the region that maps to dayside reconnection.

While this description accounts for the dominating large-scale circulation of plasma and magnetic flux in the ionosphere and magnetosphere, large variations are observed in the global morphology of ionospheric convection. Statistical studies of ground- and space-based measurements have shown that the average patterns strongly depend on the orientation of the IMF (Heppner and Maynard 1987; Weimer 2005; Ruohoniemi and Greenwald 2005; Pettigrew et al. 2010; Haaland et al. 2007). During northward IMF, the two-cell convection pattern on average reduces, and one or two small cells appear additionally at high latitudes on the dayside (Förster et al. 2008a). The convection pattern also rotates in a systematic way with changes in the IMF Geocentric Solar Magnetic (GSM) *y* component. The sense of the rotation depends on the sign of the IMF $B_y$ component, and is opposite between hemispheres. These effects can be explained in terms of different dayside magnetic field geometries (Cowley 1981), assuming that reconnection primarily occurs at the points where the IMF and the magnetosphere are most strongly anti-parallel.

The main governing mechanism behind convection takes place at high altitudes, where the Earth's field is largely dipolar, and therefore symmetrical between hemispheres. There is however evidence that the ionospheric and magnetospheric convection is modified by ionospheric conductivity (e.g., Ruohoniemi and Greenwald 2005; Pettigrew et al. 2010; Ridley et al. 2004), which has some dependence on field asymmetries (see Sect. 2). Field asymmetries have indeed been shown to modify ionospheric convection (Förster and Cnossen 2013), as does the magnitude of the Earth's dipole moment (Cnossen et al. 2011, 2012a). However, since the literature on this topic is sparse, we focus mainly on results regarding variations in the convection related to differences in conductivity. Such variations are observed both in statistical average convection patterns, which must be interpreted as representative of a quasi-steady state, and in the dynamic response of the ionosphere to changes in magnetospheric convection. We also discuss how asymmetries in the magnetic field at low altitudes introduce asymmetries in the convection when observed in a geographic reference frame.

### 3.1 Conductivity Influence on Average Convection Morphology

In the context of the present review, it is relevant to look at differences in average convection patterns related to seasonal differences. Such differences may be due to 1) effects related to reconnection geometries, which are independent of field asymmetries at low altitudes, and/or 2) effects related to ionospheric conductivity differences, which do depend on the differences summarized in Sect. 2 (e.g., Cnossen et al. 2012b). It is the latter effects that are of interest here.

When IMF $B_y$ is small, the two-cell convection pattern is not entirely symmetrical; the flow across the polar cap is slightly skewed towards dusk (e.g., Haaland et al. 2007), and the dusk cell is slightly larger. This dawn–dusk asymmetry is often attributed to ionospheric feedback associated with the Hall conductance gradients (Tanaka 2001; Lotko et al. 2014), which perturbs the magnetosphere such that the nightside reconnection region appears duskward of the Sun–Earth axis. Statistical studies of convection measurements from the Super Dual Auroral Radar Network (SuperDARN) have shown that the dawn–dusk asymmetries that appear when the IMF $B_y$ is strong are either reduced or enhanced depending on the dipole tilt angle. Ruohoniemi and Greenwald (2005) found that the asymmetries are larger for the combination $B_y > 0$/summer and $B_y < 0$/winter. They argued that these results were consistent with the Hall conductance gradient effect (Tanaka 2001). Similar results were obtained by Pettigrew et al. (2010), who also used SuperDARN measurements.

Dynamical modeling of the magnetosphere–ionosphere–thermosphere (M–I–T) interaction by Song et al. (2009) and Tu et al. (2014) has also shown that the dynamical Hall effect creates a component in the ionospheric flow which is perpendicular to the magnetospheric flow that drives it. For an anti-sunward flow, the dynamical Hall effect would create a duskward component, consistent with empirical convection patterns. The effect is stronger when the conductivity is low. The conductivity differences associated with asymmetries in the main field may therefore lead to differences in ionospheric convection even when the magnetospheric driver is symmetrical. The modeling by Tu et al. (2014) was comprehensive in the sense that it was based on a fully dynamic description of the M–I–T coupling. Their approach differs from the standard technique used in MHD models, where field-aligned currents at the ionospheric boundary are used to solve for an electrostatic potential in the ionosphere, which then is used as a boundary condition for the magnetosphere (e.g., Ridley et al. 2004). However, Tu et al. (2014) only looked at a 1D-case, solving for all electrodynamic quantities along a single vertical field line. Consequently, the dynamical Hall effect is independent of horizontal gradients in the conductivity, which are essential in the global MHD results by e.g., Lotko et al. (2014).

### 3.2 Conductivity Influence on Dynamic Response to Changes in Magnetospheric Convection

In the statistical studies of ionospheric convection cited above, the assumption has been made that **B** is static, so that the electric field is a potential field, and the magnetospheric electric field maps exactly along magnetic field lines to the ionosphere. In reality the ionosphere responds to changes in magnetospheric convection in a finite time (e.g., Song et al. 2009; Tu et al. 2014), since the magnetosphere must overcome the inertia of the ionosphere/thermosphere system before a steady state is reached. The inertia may well be different between hemispheres, due to both seasonal variations and differences in the Earth's magnetic field as discussed in Sect. 2 (see Sect. 5).

The Dungey cycle described above is not a steady circulation. Dayside and nightside reconnection tend to happen in bursts and not simultaneously, expanding and contracting the polar cap. This view is known as the expanding-contracting polar cap paradigm (Cowley and Lockwood 1992; Siscoe and Huang 1985; Milan et al. 2003; Milan 2015). In sum, the convection pattern depends both on the dayside reconnection, which can be seen as directly driven by the IMF, and on nightside reconnection. Grocott et al. (2009) showed that convection excited by nightside reconnection is much less ordered by the IMF orientation than what might be expected from the statistical studies cited above.

Each burst of reconnection is followed by a change in magnetospheric convection, to which the ionosphere takes some time to adapt. The strongest nightside reconnection events occur during substorms (Milan et al. 2007). Since substorms are also associated with a strong increase in auroral particle precipitation, the conductivity on the nightside changes dramatically. It has been shown that this conductivity enhancement is associated with a suppression of the convection (e.g., Provan et al. 2004), and that the stagnation is more prominent when the aurora is more intense (Grocott et al. 2009). The suppression is understood as an effect of enhanced friction between the charged and neutral particles as the collision frequency increases with conductivity. Indirect evidence of a seasonal difference in convection response to substorms was presented by Laundal et al. (2010a,b), who found that the substorm bulge, the footprint of newly closed field lines, was more pronounced in winter than in summer. This suggests that the ionospheric convection is more suppressed in the bulge, thus maintaining its shape, during the winter season when precipitation is on average stronger (e.g., Newell et al. 2010).

Based on the above results, it should be expected that conductivity-dependent differences in response times are also observed on the dayside. However, as far as we know, conjugate observations of the convection response to IMF changes have not provided conclusive evidence of this, as hemispheric differences in response time can also be interpreted in terms of reconnection geometry (e.g., Ambrosino et al. 2009; Chisham et al. 2000). If the conductivity does play a role in modulating ionospheric response times, we would expect a UT-dependent asymmetry in convection patterns between hemispheres due to the field asymmetries shown in Fig. 2 and pole offset differences illustrated in Fig. 4.

### 3.3 Cross Polar Cap Convection Asymmetries

The overall flux transport across the polar cap can be quantified in terms of the cross polar cap potential (CPCP), measured as the maximum electric potential difference in the polar regions. Several statistical studies have found that the CPCP is on average slightly stronger in the SH compared to the NH. Pettigrew et al. (2010), who used SuperDARN radars from both hemispheres, found a difference of 6.5%. Papitashvili and Rich (2002), who used measurements from the Defense Meteorological Satellite Program (DMSP), found a difference of 10%. Förster and Haaland (2015), who used Cluster electric field measurements mapped to the ionosphere, found differences of ∼5%–7%. All these authors cite the differences in the geomagnetic field as a possible cause for the asymmetries.

A higher CPCP in the South does not imply that the convection velocity is higher there, since the drift velocity depends on both the electric and magnetic field: $\mathbf{v} = \mathbf{E} \times \mathbf{B}/B^2$, which is proportional to $E/B$. The study by Förster and Cnossen (2013) is one of few that looks specifically at the effect of asymmetries in the field on ionospheric convection. They presented model runs, using the Coupled Magnetosphere–Ionosphere–Thermosphere model (Wiltberger et al. 2004; Wang et al. 2004), of an interval near equinox, using both a dipole field and the IGRF. They found predominantly stronger convection velocities in the NH at high latitudes (>80°) with the IGRF, and symmetrical values using the dipole. This region is representative of the cross polar cap flow. They argued that the differences could be explained by field strength asymmetries (Cnossen et al. 2011) and differences in offset between the magnetic and geographic poles (Cnossen and Richmond 2012).

Even if the CPCP is the same, the flows will be different when observed in a geographic coordinate system due to the field asymmetries. In the following we calculate mean convection velocities along the dawn–dusk meridian for an electric potential which is symmetrical between hemispheres in modified apex coordinates. We define the convection electric potential $\Phi$ such that $|\partial\Phi/\partial\lambda_m|$ is constant along the dawn dusk meridian poleward of modified apex latitude $\lambda_m = \pm 80°$ with reference height 400 km. The total CPCP is 100 kV. Using Eqs. 4.9 and 4.18 in Richmond (1995b), we calculate the corresponding drift velocity, and convert this to geographic coordinates using the software published by Emmert et al. (2010). Figure 6 (left) shows the mean convection velocity along the dawn–dusk meridian in both hemispheres as a function of UT. The maps to the right show the ±80° magnetic circles of latitude, with the dawn–dusk meridian at 00 UT in bold, and the noon meridian dashed. The diurnal variation is larger in the NH compared to the SH. The velocity in the NH peaks around 06 and 18 UT, when the orientation of the Earth is such that the major axis in the elliptical 80° contour aligns with the Sun–Earth axis. In the SH, the −80° contour is more circular, so that the diurnal variation is smaller. The interhemispheric difference in convection speed is smallest just after 00 and 12 UT, in good agreement with the modeling results by Förster and Cnossen (2013). These calculations show that even if the flux transport is the same in the two hemispheres and at all UTs, there will be a diurnal variation and a hemispheric asymmetry in convection velocities as seen in geographic coordinates.

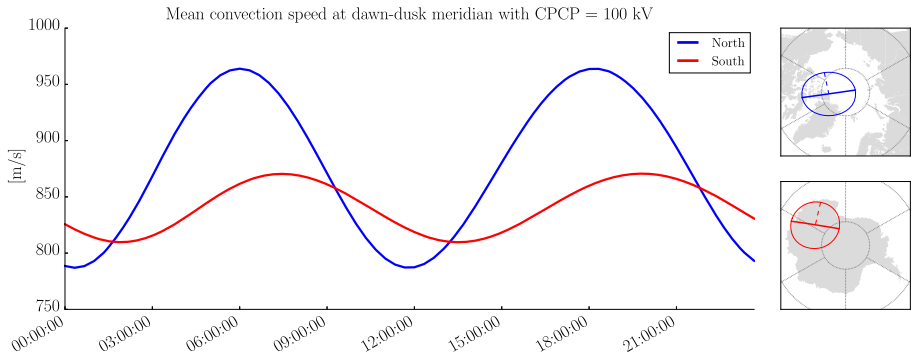

**Fig. 6** *Left*: Diurnal variation in mean convection velocities at the dawn–dusk meridian poleward of $\pm 80°$ modified apex latitude. A total CPCP of 100 kV has been used, which is constant along the dawn–dusk meridian and symmetrical between hemispheres. The panels to the right show the $\pm 80°$ modified apex magnetic latitude contours in both hemispheres, as well as the dawn–dusk meridian at 00:00 UT (*bold*) and the noon meridian at the same time (*dashed*). The calculations were done for 1 January 2015, but it is largely representative also for other times

## 4 Thermospheric Winds

### 4.1 Theoretical Considerations

High-latitude neutral winds in the thermosphere arise from a closely coupled combination of solar radiative forcing, interactions of the solar wind with the Earth's magnetosphere–ionosphere system, and ion-neutral coupling processes within the upper atmosphere. Both North–South asymmetries in plasma convection (see Sect. 3) and solar extreme ultraviolet (EUV) irradiation (see Sect. 2) can contribute to asymmetries in neutral winds, entering the momentum budget of the thermosphere via the "ion drag" and pressure gradient forces.

The ion drag force describes the momentum exchange between charged and neutral particles due to collisions between them. At high latitudes, the neutral species are usually accelerated by the (generally stronger) plasma flows driven by magnetospheric convection. The magnitude of the ion drag force depends on the difference between the ion and neutral velocities, so that North–South differences in plasma convection, as described in Sect. 3, are a first source of hemispheric asymmetry in neutral winds. However, the extent to which the ion velocities are able to influence the neutral winds also depends on the strength of the ion-neutral coupling, described by the Hall and Pedersen ion drag coefficients. In the upper thermosphere ($> \sim 150$ km) the Pedersen ion drag coefficient is much larger than the Hall ion drag coefficient and roughly proportional to the electron density (e.g., Richmond 1995a). Since solar EUV radiation plays an important role in ion and electron production, solar illumination influences the magnitude of the ion drag force to a degree.

Solar illumination is also an important factor in the pressure gradient force. Non-uniform heating due to absorption of solar EUV radiation leads to a pressure gradient directed away from the day-side equatorial region, and therefore in an anti-sunward direction across the polar region (e.g., Dickinson et al. 1981). Other processes that affect the thermospheric temperature distribution also contribute to the pressure gradient force and can modify this. At high latitudes, Joule heating is an important source of energy, especially during disturbed geomagnetic conditions. This acts to reduce the solar EUV-driven pressure gradient on the dayside, but can add to it on the nightside. The magnitude of Joule heating is dependent on both the neutral and plasma velocities, as well as the ionospheric conductivity.

Because of the role of solar radiation in both the ion drag force and pressure gradient force, differences in the amount of solar radiation received by the two hemispheres are a second source of North–South asymmetry in neutral winds. Seasonal variations associated with the tilt of the Earth's geographic axis with respect to the Sun–Earth line cause strong differences in solar illumination around solstice, when it is winter in one hemisphere and summer in the other. However, those summer–winter differences in solar illumination are not really what we are interested in here. Therefore we will compare the two hemispheres during the same local season, e.g., compare June in the NH to December in the SH. Still, even then there are differences in the average amount of illumination, as well as in spatial and diurnal variations, as explained in Sect. 2.

Hemispheric differences in the offset between the magnetic and geographic reference frames create one further source of asymmetry. Because some of the forces acting on the neutral wind are best organized in a geographic reference frame, such as the solar EUV-driven part of the pressure gradient, or the Coriolis force, while others are best organized in a magnetic reference frame, such as the ion drag force, the degree to which these two reference frames match each other influences how the different types of forcing balance and interact with each other. Consider, for example, the ion drag force and the EUV-driven pressure gradient force across the polar cap. Both are oriented in an anti-sunward direction, but the ion drag force is anti-sunward in a magnetic reference frame, while the EUV-driven pressure gradient force is anti-sunward in a geographic reference frame. The directions therefore do not match perfectly, and the discrepancy between the two is larger in the SH. In general, the greater offset between the magnetic and geographic poles in the SH leads to greater spatial differences between the two references frames and greater variations over the course of a day. These factors could therefore lead to greater variability in the SH high-latitude neutral winds, in addition to the solar illumination effect already described (see also Förster et al. 2008b; Förster and Cnossen 2013).

### 4.2 Observational and Modeling Studies

Observations made by various satellite missions and by ground-based Fabry–Pérot Interferometers (FPIs) have shown that the high-latitude neutral wind pattern exhibits a clear imprint of the ionospheric convection pattern (e.g., Thayer and Killeen 1991, 1993; Killeen et al. 1995; Emmert et al. 2006; Förster et al. 2008b, 2011) indicating the importance of the ion drag force in the thermospheric high-latitude momentum budget. For southward IMF, the neutral winds more or less follow the classic two-cell convection pattern, though with some modifications due to inertia and due to other forces acting on the neutral winds. The solar EUV-driven pressure gradient force tends to enhance anti-sunward flow across the polar cap, while inhibiting sunward return flows at lower latitudes in the dawn and dusk sectors (e.g., Thayer and Killeen 1993). Further, the neutral wind vortex on the dusk side is generally stronger than the one on the dawn side, because the Coriolis force and momentum advection term more or less balance each other on the dusk side, while they act in the same direction on the dawn side, in competition with the ion drag force (e.g., Killeen and Roble 1984; Kwak and Richmond 2007).

Förster et al. (2008b) noted systematic differences in the neutral wind patterns in the Northern and Southern polar caps, based on a statistical analysis of CHAMP data for the full year of 2003 (averaging all seasons together). In agreement with our theoretical predictions above, they found greater neutral wind variability in the SH than in the NH; standard deviations of the neutral winds in the South were about 20–40% higher than in the North. The mean neutral wind speeds in the two hemispheres were about the same (Cnossen and Förster 2015). However, further analysis of the neutral wind vertical vorticity by Förster et al. (2011) did reveal noticeable differences in magnitude. The vertical vorticity of the neutral wind isolates the rotational (non-divergent) part of the horizontal neutral wind, which is primarily associated with the ion drag force (e.g., Kwak and Richmond 2014), and is therefore expected to be more strongly influenced by plasma convection than the total wind field. The neutral wind vorticity maximum can be used as an indicator for the strength of the dawn cell, and the vorticity minimum as an indicator for the strength of the dusk cell. Using CHAMP data from two full years (2002–2003), Förster et al. (2011) showed that the magnitudes of the vorticity maximum and minimum are systematically larger in the NH, consistent with the larger ion velocities in the NH. Förster and Cnossen (2013) reproduced these North–South differences in simulations with the Coupled Magnetosphere–Ionosphere–Thermosphere (CMIT) model and demonstrated that they are associated with asymmetry in the Earth's magnetic field.

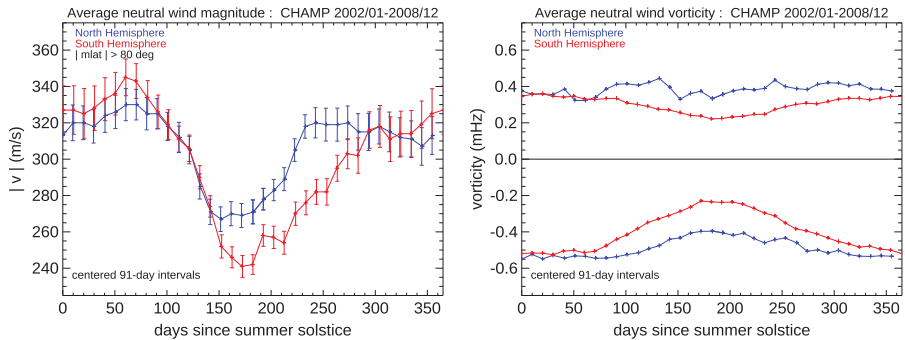

**Fig. 7** Left: 91-day averages of the mean neutral wind speed in the polar cap (>80° magnetic latitude) for the NH (*blue*) and the SH (*red*) based on CHAMP data from Jan. 2002 to Dec. 2008. *Error bars* represent the 95% confidence intervals on the means. *Right*: 91-day running averages of the maxima and minima of the high-latitude neutral wind vorticity in the same format. It was not possible to calculate the 95% confidence intervals in this case, so no error bars are shown. See Cnossen and Förster (2015) for further details

Cnossen and Förster (2015) studied the dependence of the North–South asymmetries in neutral winds on seasonal and solar cycle variations in solar illumination. A new statistical analysis of CHAMP observations from 2002–2008 showed that neutral wind speeds are always larger in the summer hemisphere, indicating the importance of solar radiative forcing on the neutral winds. However, when both hemispheres are compared for the same local season, as shown in Fig. 7, a North–South difference emerges during the winter season, with wind speeds being significantly larger in the NH. This is perhaps even clearer in the neutral wind vorticity maxima and minima, also shown in Fig. 7, suggesting that the asymmetries are forced by the North–South asymmetry in plasma convection.

The fact that the asymmetry disappears during summer might be due to North–South differences in solar radiation counter-acting the effect of the asymmetry in plasma convection. As shown in Fig. 5 the SH receives more sunlight than the NH. Since high-latitude neutral winds become notably stronger when solar irradiance is higher (e.g., Emmert et al. 2006), the larger amount of sunlight in the SH polar region opposes the effect of the larger ion velocities in the NH polar region more strongly in summer, reducing the North–South asymmetry in neutral wind speeds and vorticity, while in winter the asymmetry in solar radiative forcing is much less important.

Cnossen and Förster (2015) explored the seasonal variations in North–South asymmetry also using simulations with the CMIT model. However, the model showed generally larger neutral wind speeds and absolute vorticity values in the NH, almost regardless of the season. The model thus does not appear to reproduce the interactive balance between solar radiative effects and plasma convection effects on the neutral winds correctly, apparently placing too much emphasis on the latter. Cnossen and Förster (2015) ascribed this to a problem with the seasonal variation in electron density in the model, leading to errors in the strength of the ion-neutral coupling. The reason for the incorrect seasonal variation in electron density is still under investigation, but is likely to be complex, as the electron density distribution at high latitudes is affected by many different processes (solar EUV, energetic particle precipitation, transport by neutral winds, $\mathbf{E} \times \mathbf{B}$ drifts, etc.), which also interact with each other. This illustrates the need to better understand both the seasonal cycle and any North–South asymmetries in electron density that may be present, as discussed in Sect. 5.

## 5 Asymmetries in Total Electron Content

At F-region altitudes, production and loss of ions and electrons are governed by solar EUV radiation along with the thermosphere composition. In particular, photoionization of atomic oxygen (O) is the primary source of $O^+$, which dominates the F-region plasma population. The loss of $O^+$ is due to ion exchange reactions with molecular nitrogen ($N_2$) and molecular oxygen ($O_2$). Spatial and temporal variability in either the EUV radiation or thermosphere composition will therefore have a direct impact on the F-region electron density. Though not discussed here, neutral winds and ionosphere electric fields additionally contribute to the ionosphere variability through the redistribution of plasma to regions of increased or decreased production and loss.

Asymmetries between the geomagnetic field in the Northern and Southern Hemispheres (Sect. 2) introduce an asymmetry in the solar EUV radiation and the neutral composition, leading to hemispheric differences in the F-region electron density. This is primarily due to the offset between the magnetic and geographic poles. As shown in Figs. 4 and 5, there are considerable differences in the solar illumination of high magnetic latitudes in the Northern and Southern Hemispheres. North–South asymmetry in the magnetic field, together with variations in the Sun–Earth distance, result in the SH high latitude ionosphere experiencing greater exposure to EUV radiation compared to the NH. Additionally, energy inputs at high latitudes and changes in the (horizontal and vertical) transport modifies [$O/N_2$]. As the energy input is related to the geomagnetic field geometry, the thermosphere composition, and its impact on production and loss of ions and electrons, will be impacted by hemispheric asymmetries in the geomagnetic field.

To illustrate the differences between the ionospheres in the Northern and Southern Hemispheres, Figs. 8 and 9 show the nighttime (00 MLT) and daytime (12 MLT) total electron content (TEC) from the Constellation Observing System for Meteorology, Ionosphere, and Climate (COSMIC) Global Positioning System (GPS) radio occultation observations (Anthes et al. 2008) under equinox, winter, and summer conditions. The COSMIC TEC observations are the integrated electron density up to ∼800 km, and are thus dominated by the electron density at F-region altitudes. Note that the results in Figs. 8 and 9 are presented in terms of magnetic apex latitude and longitude, and are based on geomagnetic quiet (Kp < 3) observations during the solar minimum years of 2007–2009. Differences in the NH and SH TEC are clearly apparent. First, one can see that during March equinox, the average TEC poleward of 60° is slightly larger in the SH compared to the NH. Notably larger values of TEC also occur in the SH during local summer (i.e., December solstice in the SH and June solstice in the NH). However, during local winter, the NH TEC is greater than the SH

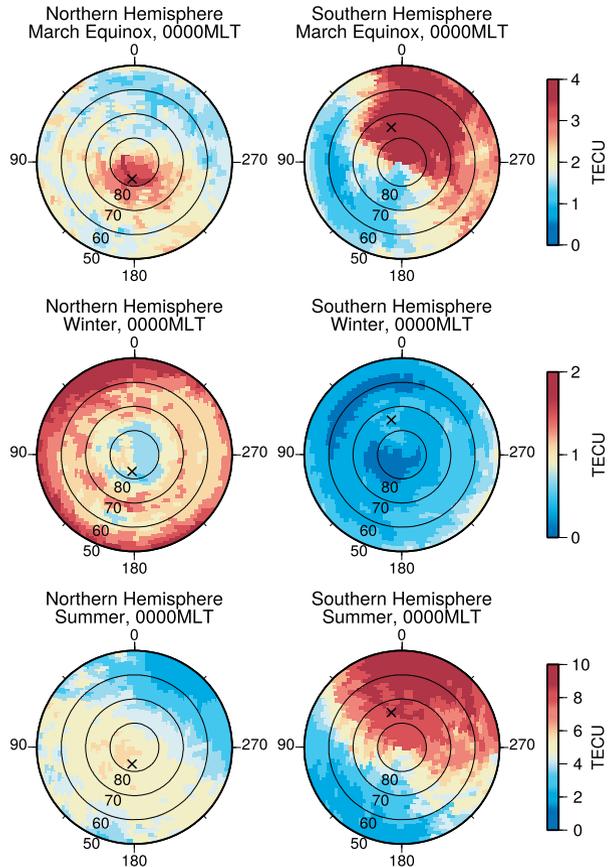

**Fig. 8** COSMIC TEC at 0000 MLT in the NH (*left panels*) and SH (*right panels*) for March equinox (*top panels*), local winter solstice (*middle panels*), and local summer solstice (*bottom panels*). Results are presented in Apex latitude and longitude, and are the average of geomagnetically quiet days for 2007–2009. The geographic pole positions are marked by crosses

TEC. The 2007–2009 average F10.7 cm solar flux during December and June solstice is nearly equivalent, and the hemispheric differences during local winter and summer are thus unrelated to changes in solar activity between December and June solstices.

Differences in the longitudinal distribution of the TEC between the two hemispheres are also evident in Figs. 8 and 9. In particular, during the daytime (12 MLT, shown in Fig. 9), the TEC is preferentially larger in the magnetic apex longitude sectors that are furthest from the geographic pole, which is marked by crosses. Longitude sectors far from the geographic pole are most sunlit, and thus the daytime TEC is larger near 0° apex longitude in the NH and 180° longitude in the SH. The opposite occurs during the nighttime (00 MLT, shown in Fig. 8), when the TEC is greater at longitudes which are closer to the geographic pole.

The hemispheric asymmetries that are present in Figs. 8 and 9 can largely be explained by seasonal variations in the Sun–Earth distance, and hemispheric differences in the geomagnetic field. The change in solar radiation due to varying Sun–Earth distance between the December and June solstices (see Fig. 5) results in greater winter electron densities in the NH and larger electron densities in summer in the SH. This leads to an ∼7% difference (Zeng et al. 2008), and explains a portion of the hemispheric asymmetry during solstice conditions in Figs. 8 and 9. Variations in the Sun–Earth distance cannot, however, explain the relatively larger TEC that occurs in the SH during March equinox. During March equinox, the average nighttime (Fig. 8, upper panels) TEC is similar in both hemispheres, and is dom-inated by longitudinal variations that arise due to thermosphere composition, which will be discussed later. During daytime (Fig. 9, upper panels), we attribute the larger TEC at March equinox in the SH to the different offset between the geographic and magnetic poles in the two hemispheres, which leads to solar EUV radiation occurring at higher magnetic latitudes in the SH. We note that this mechanism also impacts the results during solstice time periods; however, it is less evident in Figs. 8 and 9 due to the aforementioned impact of variations in Earth–Sun distance. As explained by Zeng et al. (2008), the tilt of the geomagnetic field also drives differences in the longitudinal variations in the Northern and Southern Hemispheres. In particular, the tilt of the geomagnetic field leads to magnetic longitudes further from the geographic poles receiving more solar EUV radiation during the daytime, resulting in greater daytime TEC at these longitudes (Fig. 9). When solar EUV forcing is largely absent, variations in thermosphere composition are thought to be responsible for the different longitudinal variability in the Northern and Southern Hemispheres. Regions of enhanced downwelling, which increases the [O/$N_2$] ratio, tend to occur in the magnetic longitude sector of the geographic pole (e.g., Rishbeth and Müller-Wodarg 1999), resulting in the observed enhancement in night time TEC in longitude sectors near the geographic pole compared to regions further away (Fig. 8). The longitudinal variations in the [O/$N_2$] ratio are driven by longitudinal variations in thermosphere circulation, which arise due to the influence of the geomagnetic field on the spatial distribution of the high latitude energy input.

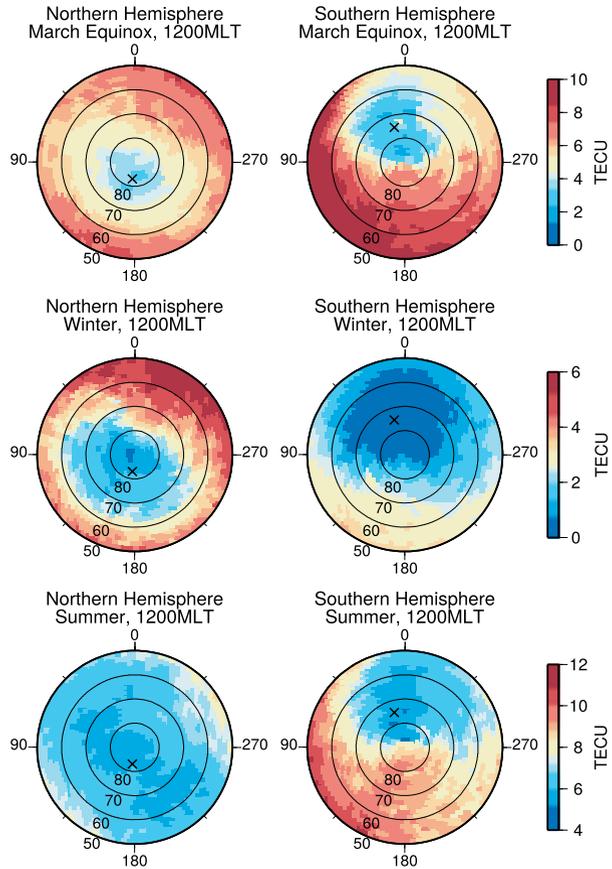

**Fig. 9** Same as Fig. 8, except for the results are shown for 1200 MLT

## 6 Asymmetries in Ion Outflow

The thermosphere continually loses matter in the form of ion outflow. Estimated loss rates are about $10^{26}$ ions/sec from both hemispheres combined (e.g., Yau and Andre 1997). Although observations pointing out North–South asymmetries in ion upflow and ion outflow exists (e.g., Zhao et al. 2014, and references therein), this issue has not been extensively addressed. Model and simulation results are also scarce, but a recent study by Barakat et al. (2015) demonstrated that North–South asymmetries in outflow are reproduced if realistic boundary conditions are used to parametrize models. To our knowledge, North–South asymmetries are not explicitly built into large scale models of the magnetosphere either.

When discussing ion outflow, it is natural to divide the source areas of ionospheric outflow into two distinct regions, the auroral zone and the cusp region on one side and the high latitude open polar cap on the other side. Processes and characteristics of the outflow are very different between these regions, but there can be significant horizontal transport of plasma between regions.

Two fundamental elements are necessary for ion outflow: first, ionization, which provides a source of free ions, and second, acceleration processes able to give the ions sufficient energy to escape the Earth's gravitational potential. For the most relevant species for Earth, $H^+$ and $O^+$, escape energies are of the order of 0.6 and 10 eV, respectively. North–South asymmetries can exist in both ionization and transport.

### 6.1 Auroral Zone and the Cusp Region

On a large scale, the nightside auroral zone is characterized by enhanced outflow which largely balances the electron precipitation responsible for auroral arcs. Ionization, at least on the nightside where EUV illumination is absent, is primarily driven by the auroral precipitation (e.g Hultqvist et al. 1999). The outflow is mainly driven by strong field aligned electric fields caused by anomalous resistivity, and both $H^+$ and $O^+$ can be extracted and accelerated to escape energies. Furthermore, the nightside auroral zone is co-located with a region of Birkeland (magnetic field-aligned) currents and strong flow shears which locally tend to break up into vortices. Such small scale structures may provide an additional source of energy for plasma escape.

Except for the study by Zhao et al. (2014), based on measurements from the Fast Auroral Snapshot (FAST) satellite of $H^+$ in the 1 eV–1.2 keV energy range for the years 2000–2005, no systematic studies trying to quantify North–South asymmetries in ion outflow from the auroral zone exists. Since the outflow is highly correlated with precipitation, however, much of the asymmetries related to aurora, discussed in Sect. 8 are also relevant for ion outflow from this region. Processes responsible for outflow from the cusp region are to some extent comparable to those of the auroral zone; ionization occurs partly by sunlight and partly by electromagnetic energy.

### 6.2 The Polar Cap

Poleward of the auroral zone, in the polar cap regions, there is little or no significant precipitation, and consequently no electric field set up by anomalous resistivity. The outflow seems to be limited by ionization (André et al. 2015; Kitamura et al. 2015), and since ionization is largely driven by EUV illumination, there are diurnal and seasonal variations and thus an inherent North–South asymmetry. Observations of such asymmetries have been reported by e.g., Kitamura et al. (2015) and are also corroborated by model results, e.g., Glocer et al. (2012)

The energy required to escape the gravitational potential comes from a combination of thermal forces and an ambient electric field set up by charge exchange. The available energy is lower than in the cusp and auroral zone, so outflow from the polar cap region is dominated by cold (energies up to a few 10's of eV) protons. In addition to the ambient electric field, mirror forces and centrifugal acceleration can also provide parallel acceleration. The mirror force depends on the magnetic field, and thus possesses a North–South asymmetry (see Sect. 2). Likewise, the centrifugal acceleration is governed by the convection, which may be North–South asymmetric (See Sect. 3).

Cold ions are notoriously difficult to measure in-situ, and have often been termed invisible (e.g., Chappell et al. 1987, 2000; André and Cully 2012). Their low energy combined with shielding effects due to spacecraft charging issues usually prevents detection with particle instruments, so alternative methods are needed. The first large scale survey of cold ions (Engwall et al. 2009) was based on observations from the Cluster mission and a wake detection technique (Engwall et al. 2006). North–South asymmetries in cold outflow were reported by Li et al. (2012), but due to the orbit of Cluster, a quantitative assessment of the asymmetry is difficult.

A recent simulation study by Barakat et al. (2015) discusses effects of the difference in magnetic pole offset between the two hemispheres (see Sect. 2.2) and its consequence for ionospheric outflow. Their simulation results are for a geomagnetic storm around equinox, and show larger diurnal modulation in the southern hemisphere. They attribute the North–South asymmetry to the offset difference, and suggest that the hemispherical asymmetry and periodicity of the total ion outflow could influence the magnetospheric tail and perhaps contribute to substorm triggering.

In addition to local ionization in the polar cap region, upwelling $O^+$ ions near the cleft can form an ion fountain (Lockwood et al. 1985) where the upwelling ions can be transported into the polar cap by anti-sunward convection.

## 7 Asymmetry in Ionospheric Currents and Magnetic Field Perturbations

In Sect. 2 we showed that the asymmetries in the Earth's magnetic field lead to differences in ionospheric conductivity, due to 1) a dependence on the field strength in the sunlight induced conductances (Richmond 1995a) and 2) differences in offset between magnetic and geographic poles, which lead to differences in diurnal variation in sunlight exposure in the polar region, which have large implications for the conductivity (Robinson and Vondrak 1984; Moen and Brekke 1993).

The differences in conductivity between hemispheres naturally have implications for differences in ionospheric currents and associated magnetic field perturbations. The relationship between the Hall and Pedersen conductance and the Hall, Pedersen and Birkeland (field-aligned) currents can be described in terms of the ionospheric Ohm's law. The horizontal part of Ohm's law is

$$\mathbf{J}_\perp = \Sigma_H \mathbf{B} \times \mathbf{E}/B + \Sigma_P \mathbf{E}, \qquad (1)$$

where we have made the idealized assumption of zero neutral wind. $\mathbf{E}$ is the electric field which appears because any large-scale electric field in the reference frame of the plasma is zero. It is therefore related to the plasma velocity (see Sect. 3) by $\mathbf{E} = -\mathbf{v} \times \mathbf{B}$. The divergence of this equation, assuming current continuity, gives the Birkeland current:

$$j_\parallel = \Sigma_P \nabla \cdot \mathbf{E} + \mathbf{E} \cdot \nabla \Sigma_P + (\mathbf{E} \times \nabla \Sigma_H) \cdot \mathbf{B}/B \qquad (2)$$

where the $\nabla$ operators act only horizontally. It is clear that the current magnitudes are highly dependent on conductivity. The Hall current scales with the Hall conductance, and the Pedersen current with the Pedersen conductance. The Birkeland currents are most strongly dependent on the Pedersen conductance.

Ground magnetometers sense only what is called an equivalent current, which is not necessarily equal to any of the current components described above. At high latitudes, the equivalent currents are equal to the divergence-free component of the horizontal ionospheric currents (e.g. Fukushima 1994; Vasyliunas 2007, and references therein). Which part of the actual current system constitutes the divergence-free horizontal currents depends on the conductivity. When the conductance gradients are zero, or perpendicular to electric equipotential contours, the equivalent current is equal to the Hall current. Laundal et al. (2015) showed that during sunlit conditions in the polar cap, the equivalent current typically aligns with the overhead Hall current. In dark conditions, the equivalent current tends to align with an overhead current which is anti-parallel to the horizontal closure of the Birkeland current system. This is consistent with the actual current being approximately zero in the polar cap in darkness. It is also consistent with observed differences in disturbance field morphology between different seasons (Friis-Christensen and Wilhjelm 1975).

Both ionospheric currents and the associated magnetic disturbances depend on quantities that are best organized in different coordinate systems: The ionospheric convection (and $\mathbf{E}$), as well as the conductance produced by auroral precipitation, are organized in magnetic coordinates, while the component of the conductances that is produced by solar EUV flux is best organized in geographic coordinates. Therefore the distribution of sunlight on magnetic apex/CGM grids in the two hemispheres is never symmetrical, and perfect hemispheric symmetry in the current and magnetic disturbance fields can not be expected either.

To illustrate this point we look at the seasonal and diurnal variation in magnetic field perturbations at two pairs of nearly conjugate magnetometers. Their locations are indicated in the top left map in Fig. 4: The filled circles show the positions of the UMQ station (at 75.6° apex latitude, and 41.2° longitude in 2015) in blue and the B22 station (at $-75.7°$ and 30.8°) in red. The triangles mark the LYR station (at 75.4° and 109.2°) in blue and the DVS station (at $-74.7°$ and 102.3°) in red. They are all at nearly the same magnetic latitude, but their locations relative to the geographic poles are different. Figure 10 shows the mean magnetic perturbation at these magnetometer stations as a function of universal time hour and month. The SuperMAG baseline subtraction has been used, which is designed such that the remaining signal can be interpreted as being associated with external (solar wind/magnetospheric) drivers (Gjerloev 2012). Diurnal variations associated with the solar quiet (Sq) currents are removed. Conjugate pairs are shown in the same columns.

We see that the seasonal variation at the conjugate stations is approximately in antiphase, due to the hemispheric difference in sunlight illumination. The contours mark the time when the mean solar zenith angle is 90°, i.e., the demarcation between the magnetometer being predominantly sunlit or not. The largest average magnetic perturbations occur at times when the magnetometer was sunlit. Comparing the two magnetometers in the SH, we see that there is most often a stronger diurnal variation at the DVS station compared to B22. This can be understood as an effect of the B22 station being much closer to the geographic pole ($-86.5°$ geographic latitude) compared to DVS ($-68.6°$), and thus experiencing less variation in sunlight during a day. Hence the more horizontal sunlight terminator contours at this location. In the winter months, when both stations are in darkness, the diurnal variation has a similar magnitude at B22 and DVS.

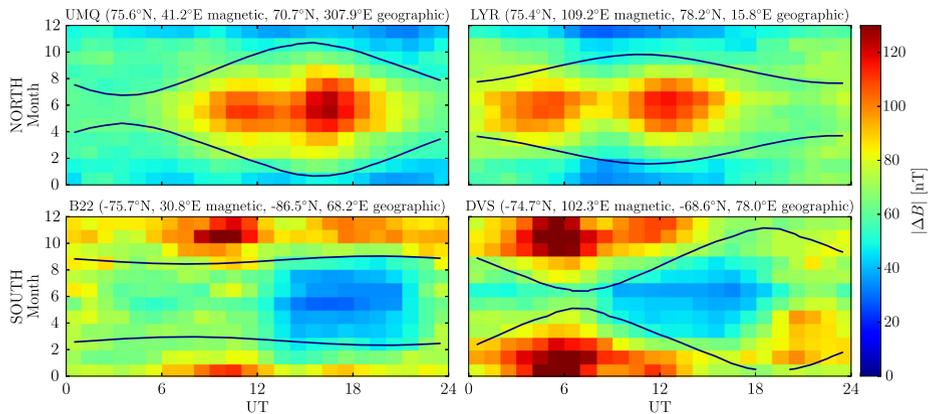

**Fig. 10** Mean magnetic field perturbations as a function of UT hour and month at two nearly conjugate magnetometer pairs close to $\pm 75°$ apex latitude. Contours of mean solar zenith angle $90°$ are also shown, indicating the demarcation between sunlight and darkness at the different stations

It is worth noting that at certain UTs, the difference between sunlit and dark conditions is modest, and in some cases even opposite to the general picture (e.g. at 20–23 UT at the DVS station). This indicates that solar illumination may be less important in certain magnetic local times. At 20–23 UT, the DVS and LYR stations are close to magnetic midnight. Being at relatively high latitudes, often inside the polar cap, the magnetic perturbations at these times may be associated with substorm poleward expansions, during which intense precipitation enhances the conductivity.

A number of previous studies have also investigated hemispheric differences and similarities in ground magnetometer measurements (see review by Wescott (1966) and the work by e.g., Viljanen and Tanskanen (2013), Weygand et al. (2014) and references therein). Most of these studies have looked at time series, showing largely similar perturbations in the two hemispheres, which indicates that changes in ionospheric convection, and consequently currents, most often occur simultaneously in the two hemispheres (Yeoman et al. 1993). Hajkowicz (2006) found a seasonal variation in the level of correspondence between time series in conjugate magnetometers, consistent with a conductivity effect.

It has also been shown that the auroral electrojet indices (AE) exhibit a UT variation which varies with seasons (Ahn et al. 2000; Singh et al. 2013). This variation is probably due to variations in conductivity, as well as the non-uniform magnetometer coverage used to derive the indices. Laundal and Gjerloev (2014) repeated the study by Singh et al. (2013), using apex quasi-dipole magnetic field components instead of the standard $H$ component (or in this case, the SuperMAG $N$ component, which is similar to $H$). A significant fraction of the UT variation was removed by this change, which indicates that the longitudinal variation in the Earth's magnetic field is contained in the UT variation of the traditional AE indices (Gasda and Richmond 1998). These studies were based on magnetometer stations in the NH. Since the longitudinal variation is different in the SH, and since the conductivity is different, an AE index derived from SH magnetometer measurements would be different from the standard index, even if the magnetometers were at conjugate points to those in the North. This was indeed shown by Weygand et al. (2014), who used SH magnetometers that were close to the conjugate points of the NH AE stations.

The effect of magnetic field strength on conductance produced by sunlight (Richmond 1995a; Cnossen et al. 2011, 2012a) has to our knowledge not been directly detected in studies of high latitude magnetic perturbations. The effect could of course be implicit in the results showing longitudinal and hemispheric variations, which most often is explained in terms of pole offsets.

### 7.1 Asymmetries in the Birkeland Currents

The asymmetries in the ionospheric conductivity (Sect. 2) also lead to asymmetries in the Birkeland (field-aligned) currents, which electrodynamically link the ionosphere to the magnetopause and the partial ring current. Studies have shown that the Birkeland currents increase in intensity during the summer (Fujii et al. 1981; Ohtani et al. 2005), and measurements of the Birkeland currents have been used to quantify variations in the ionospheric conductivity with solar zenith angle (Fujii and Iijima 1987).

Later studies have shown that the currents also exhibit a hemispherical asymmetry in MHD modeling (Wiltberger et al. 2009). However, investigations of vorticity in the ionospheric convection have shown increased vorticity during summer, which may imply that the hemispherical asymmetry in the Birkeland currents is not wholly due to variations in conductance (Chisham et al. 2009). Some authors have suggested that only the dayside currents become larger during the summer (Wang 2005), such that the hemispherical asymmetry is limited to currents on the dayside.

More recently, Coxon et al. (2015) conducted a study of Birkeland currents measured by the Active Magnetosphere and Planetary Electrodynamics Response Experiment (AMPERE) which showed that seasonal and diurnal variations in current magnitude in the Northern and Southern Hemispheres were consistent with changes in solar insolation. Figure 11 shows monthly averaged Birkeland currents in the two hemispheres for the 36 months of 2010 to 2012. In the NH there is a clear seasonal variation, with current magnitudes peaking around NH summer months (red shading). For reasons that will be discussed below, the seasonal variation in the SH current magnitudes, which are expected to maximize in SH summer months (blue shading), is less pronounced.

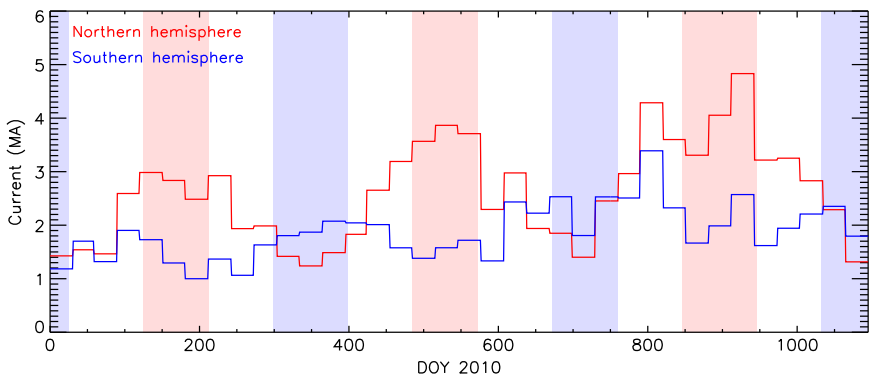

**Fig. 11** The monthly averaged Birkeland currents from January 2010 to December 2012. The NH is shown in *red* and the SH is shown in *blue*; *pink* and *light blue* shading show summer in the Northern and Southern Hemispheres respectively. Adapted from Coxon et al. (2015)

Another variation was also discovered, in which both Northern and Southern currents varied in sync: for instance see the similarity in behaviors in the two hemispheres between days 800 and 950. Such variations are associated with changes in the monthly averaged strength of solar wind-magnetosphere coupling, dependent on conditions in the solar wind. When this is corrected for, using a model developed by Milan (2013), the expected seasonal variations in the two hemispheres become readily apparent. The lack of a clear seasonal variation in the winter hemisphere was due to a coincidental antiphase between changing solar wind conditions and SH conductance levels.

Coxon et al. (2015) concluded that solar wind-magnetosphere coupling drives magnetosphere–ionosphere coupling currents in each hemisphere, but that the magnitude of these currents depends on the seasonal variation in conductance in each polar ionosphere. One last puzzle remains, however. Even when the solar wind variations are accounted for, the current magnitudes in the NH are on average greater than the currents in the SH (as is apparent in Fig. 11). It is not yet clear if this is a real effect or an artifact of the AMPERE analysis technique.

## 8 Asymmetry in the Aurora

It is well established from statistical studies (Shue et al. 2001; Coumans et al. 2004; Newell et al. 2010; Reistad et al. 2014) and from studies of conjugate images (Ohtani et al. 2009; Laundal and Østgaard 2009; Reistad et al. 2013; Fillingim et al. 2005; Stenbaek-Nielsen and Otto 1997; Sato et al. 1998) that the intensity of the aurora and the characteristics of particle precipitation can be quite different at conjugate points. These differences are mainly related to seasonal variations, and to asymmetric solar wind forcing on the magnetosphere, when the IMF has a significant GSM $y$ (and to a lesser degree $x$) component. The IMF effect on the aurora is presumably independent of differences in the main field, since the Earth's field is largely a dipole at the altitudes where the solar wind-magnetosphere interaction happens. The seasonal differences can likely be attributed to the orientation of the dipole axis with respect to the Sun–Earth line, and to variations in ionospheric conductivity. The latter will vary between hemispheres as described in Sect. 2.

The most comprehensive statistical study of the seasonal variation of particle precipitation was done by Newell et al. (2010), who analyzed a large set of particle spectra measured by instruments on the Defense Meteorological Satellite Program (DMSP) satellites. They analyzed seasonal variations in the electron and ion energy flux and number flux for three different types of precipitation, characterized by the spectrum: Monoenergetic, broadband (only electrons) and diffuse precipitation (both ions and electrons). Monoenergetic electron precipitation is believed to be accelerated by parallel electric fields, while broadband precipitation is accelerated by Alfvén waves. Diffuse precipitation, which makes up most of the energy flux (Newell et al. 2009), consists of particles that are scattered into the loss cone and not necessarily accelerated further. All types of electron aurora were found to be stronger on the nightside during winter. The winter/summer ratio was much stronger for monoenergetic precipitation (1.70) compared to broadband (1.26) and diffuse (1.30) precipitation. On the dayside however, the winter/summer ratio was less than 1 for all types of aurora except diffuse electron aurora during strong solar wind driving. The strong seasonal differences on the nightside might be explained by a feedback mechanism (Lysak 1991), by which increased ionization from precipitation leads to stronger currents and more precipitation (Ohtani et al. 2009). The differences on the dayside, which are in an opposite sense compared to the nightside, may be explained by a combination of 1) a more favorable geometry during summer for direct ion entry from the magnetopause to the ionosphere in the cusp, and 2) stronger field-aligned currents on the dayside in the summer (e.g. Green et al. 2009). The latter effect is likely to depend on the conductivity at the ionospheric footpoints, which varies differently in the two hemispheres due to asymmetries in the Earth's magnetic field (Fig. 4).

Much less is known about the importance of differences in field strength at conjugate footpoints. The most comprehensive study of this effect so far was based on data from a series of 18 conjugate flights carrying calibrated all-sky cameras along the magnetic meridian at College, Alaska between 1968 and 1971 (Belon et al. 1969; Stenbaek-Nielsen et al. 1973). These data showed that the aurora was brighter, more frequent, and more extended in latitude in the NH, where the magnetic field was weakest. During very active times, and at the highest latitudes, the differences were less systematic.

To explain and quantify the magnetic field control on auroral intensity, Stenbaek-Nielsen et al. (1973) developed a model for three idealized cases of pitch angle distribution, corresponding to different degrees of scattering, and also allowing for parallel electric potentials which may be different in the two hemispheres. The quantities derived in their paper were representative of the magnetic field differences at College, Alaska and the conjugate hemisphere. Here we briefly review their model, and present global maps of the expected inter-hemispheric differences for the different pitch angle distributions.

Conservation of the first adiabatic invariant, $mv_\perp^2/2B$, implies that the relationship between the pitch angle of a particle when it crosses the equatorial plane, $\alpha_{eq}$, the equatorial magnetic field strength along its trajectory, $B_{eq}$, and the magnetic field strength at which the particle mirrors, $B_m$, is:

$$\sin^2 \alpha_{eq} = B_{eq}/B_m. \tag{3}$$

Assuming that all particles that mirror below some fixed height precipitate, and those that mirror above this height escape back into the magnetosphere, the destiny of a particle can be determined by its pitch angle in the equatorial plane. Particles that have pitch angles less than a certain limit are within the loss cone, given by

$$\alpha_l \approx \sqrt{B_{eq}/B_m}. \tag{4}$$

We have used that $\sin \alpha \approx \alpha$, since the ratio in Eq. 3 is always small for particles that mirror at ionospheric altitudes.

Consider an equatorial cross section of a flux tube with area $A_{eq}$. Assuming an isotropic pitch angle distribution, the number flux of particles through this cross section that eventually precipitate can be expressed in terms of the directional particle flux, $j$, times the area and the solid angle of the loss cone:

$$n \approx A_{eq} \pi j \alpha_l^2, \tag{5}$$

where the small loss cone angle assumption has been used.

Since the magnetic field strength may be different at conjugate points, the loss cone may be different for the two hemispheres. The ratio between the loss cones in the two hemispheres can be written (using Eq. 4):

$$R_{\alpha_l} = \alpha_l^n/\alpha_l^s = \sqrt{B_m^s/B_m^n}, \tag{6}$$

where the superscripts denote the hemisphere. These equations imply that the number of particles precipitating to each hemisphere is different when $B_m^s \neq B_m^n$. However, the area of the flux tube at the two mirror points will also be different in that case, and this effect may balance the number flux when considering the number of particles per area (the intensity). Whether or not that happens depends on the pitch angle distribution and the geometry of the light. Stenbaek-Nielsen et al. (1973) considered three different pitch angle distributions:

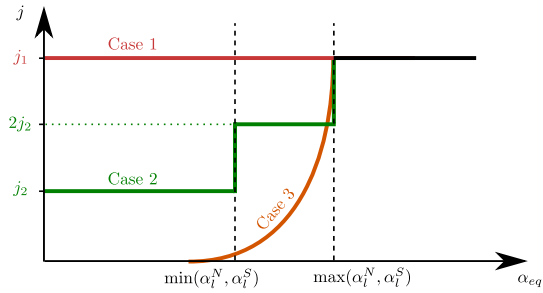

**Fig. 12** The three idealized pitch angle distributions considered by Stenbaek-Nielsen et al. (1973) in order to estimate the inter-hemispheric asymmetries in particle precipitation and auroral luminosity due to differences in field strength. The *vertical dashed bars* denote the loss cones in the equatorial plane, which may be different in the two hemispheres. The distribution outside the loss cones is isotropic, at the level shown by the *horizontal black line*. See text for details

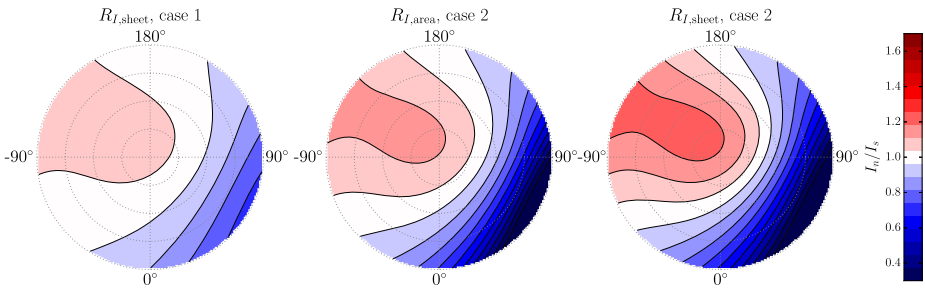

**Fig. 13** Intensity ratios at conjugate points for two of the three cases of pitch angle distributions, considered by Stenbaek-Nielsen et al. (1973) and illustrated in Fig. 12. For case 1, only the ratios for aurora which appears in a thin sheet is shown. For the case that the aurora is spread over a large area, the ratio would be 1 everywhere. The maps for case 2 correspond to aurora distributed over an area (*middle*) and aurora which appears in a thin sheet (*right*). Case 3 is not shown. See text for details

**Case 1: Isotropic Distribution, Strong Diffusion** The particles are strongly scattered, so that the loss cone is constantly refilled at a rate which balances the loss to precipitation. The pitch angle distribution is isotropic (See Fig. 12). In this case, the intensity ratio becomes

$$R_I = \frac{n^n A_m^s}{A_m^n n^s} = 1, \qquad (7)$$

where we have used Eqs. 3 and 5, and magnetic flux conservation, $B_m^n A_m^n = B_m^s A_m^s$. The area differences balance the difference in particle flux. However, if the aurora appears in a thin sheet, it can be considered a two-dimensional structure. Then it might be more relevant to consider the number of particles per unit length rather than area. Since the length scales as the square root of the area, we get the intensity ratio:

$$R_{I,\text{sheet}} = \frac{n^n \sqrt{A_m^s}}{\sqrt{A_m^n} n^s} = \sqrt{\frac{B_m^s}{B_m^n}} \qquad (8)$$

which means that the intensity of thin auroral sheets may be different if the field strength is different. A map of the ratio $\sqrt{B_m^s/B_m^n}$ is shown in Fig. 13.

**Case 2: Anisotropic Distribution, Strong Diffusion** The particles are strongly scattered, and the loss cone is refilled at an equal rate everywhere, but not fast enough to balance the loss to the atmosphere. This results in a step-like pitch angle distribution. Since the loss of particles is approximately twice as fast inside both loss cones than it is at $\alpha$ between $\min(\alpha_l^n, \alpha_l^s)$ and $\max(\alpha_l^n, \alpha_l^s)$, there will be a factor of 2 difference between the fluxes in these regions (see Fig. 12).

Since the flux is isotropic within both these regions, we can use Eq. 5 to get the ratio between the number of particles that precipitate per unit area to the two hemispheres per unit time. For the case that the field is strongest in the SH, the number flux unit area becomes:

$$R_I = \frac{n^n A_m^s}{A_m^n n^s} \approx \frac{A_m^s}{A_m^n} \frac{A_{eq} \pi j \alpha_l^{s\,2} + 2 A_{eq} \pi j (\alpha_l^{n\,2} - \alpha_l^{s\,2})}{A_{eq} \pi j \alpha_l^{s\,2}}$$
$$= \left(2\frac{B_m^s}{B_m^n} - 1\right) \frac{A_m^s}{A_m^n} = 2 - \frac{B_m^n}{B_m^s}. \tag{9}$$

To get the number flux per length (intensity in the case of 1-dimensional, sheet-like aurora), we use the square root of the area fraction in the last line:

$$R_{I,\text{sheet}} = \left(2\frac{B_m^s}{B_m^n} - 1\right)\sqrt{\frac{A_m^s}{A_m^n}} = \left(2 - \frac{B_m^n}{B_m^s}\right)\sqrt{\frac{B_m^s}{B_m^n}}. \tag{10}$$

Since $B_m^s > B_m^n$ in these equations, we see that the asymmetry in intensity is larger for sheet-like auroras than for aurora which is distributed over a larger area. When the SH field is weaker than in the NH, the equations above must be changed accordingly. Maps of $R_I$ and $R_{I,\text{sheet}}$ for the case of strong scattering and anisotropic pitch angle distribution are also shown in Fig. 13.

**Case 3: Weak Diffusion** The particles are only weakly scattered, and the time it takes to refill the loss cone is larger than the bounce time. The particles predominantly precipitate to the hemisphere with the weakest magnetic field (see Fig. 12). In this case, the ratio between the intensities in the two hemispheres can be infinite.

The above ratios can be modified by any net difference in field-aligned electric potential. In-situ measurements of particle precipitation accelerated by parallel electric fields have shown that the electric fields are stronger and more frequent in darkness (Newell et al. 1996, 2010). Therefore net potential drops between hemispheres almost certainly exist, and particularly during solstices. Stenbaek-Nielsen et al. (1973) showed that a net inter-hemispheric potential difference will lower or raise the mirror point, such that the intensity ratios above, $R_I$, are scaled by a factor of $1 + 2\Delta W/W_{eq}$, where $\Delta W$ is the energy difference introduced by the net potential difference (positive when the NH is at higher potential), and $W_{eq}$ is the energy of the particles in the equatorial plane. A consequence of the dependence on $W_{eq}$ is that the inter-hemispheric differences should be more pronounced for less energetic particles.

The measurements from the conjugate flight campaigns arguably still remain the strongest observational evidence of a relationship between the auroral intensity and the strength of the Earth's magnetic field. Frank and Sigwarth (2003) reported observations from one single event of the aurora in both hemispheres observed from the Visible Imaging Earth camera on board the Polar spacecraft, which was positioned such that both auroral regions were visible. They found that the aurora was brighter in the NH compared to the SH by tens of percent. This observation was made at $\approx -150°$ magnetic longitude, where the NH field is weaker. Thus it is consistent with the explanation in terms of field asymmetry described above. Note that in their manuscript they get the field asymmetry at the location of the observations wrong, but they also get the mechanism by which field asymmetries work wrong, resulting in the right conclusion with respect to the Stenbaek-Nielsen et al. (1973) model.

If the mechanism outlined above is important for the overall intensity of the aurora, it can be expected that this is also reflected in the longitudinal variation of its intensity. A few studies have looked at the longitudinal variation. Stenbaek-Nielsen (1974) analyzed data from ground all-sky imagers from the international geophysical year, when a substantial number of such cameras were operated. They found that the occurrence rate of aurora varied with longitude in a similar manner as the inter-hemispheric difference in magnetic field strength at conjugate points at 65° latitude. They interpreted this as indication that the magnetic field strength also controls discrete aurora, which was what the all-sky cameras primarily observed. Indirect evidence of a similar longitudinal variation was presented by Barth et al. (2002), who looked at observations of NO, produced by electron precipitation. The NO distribution had a similar longitudinal variation as the field differences at conjugate points.

The first study which included the longitudinal variation in auroral energy flux from both hemispheres was conducted by Luan et al. (2011). They analyzed auroral power in the 21-03 MLT sector, based on data from the global ultraviolet imager (GUVI) onboard the Thermosphere Ionosphere Mesosphere Energetics and Dynamics (TIMED) spacecraft from between 2002 and 2007. They did find a similar longitudinal pattern as those reported by Stenbaek-Nielsen and Barth et al., at least during summer and equinox seasons. Surprisingly however, they also found largely the same longitudinal variation in the opposite hemisphere. This is not expected if field asymmetries control the longitudinal variation, since this would produce an opposite pattern in the conjugate hemisphere. They found that the peak intensities coincide with the longitudes at which there is least sunlight. During local winters, when the 21-03 MLT region is always in darkness, they found that the peak intensities coincide with the darkest longitudes in the opposite hemisphere. The correlations with field strength were small, suggesting that this only plays a minor role in generating longitudinal variations in auroral intensity.

Based on Luan et al.'s work, it seems that sunlight, and consequently the effect of differences in the alignment between geographic and magnetic coordinates, is a more important factor than field strength in controlling the distribution of auroral intensity. However, more simultaneous measurements of the aurora in the two hemispheres are needed to draw firm conclusions about this. So far, the only truly comparable instrumentation providing such measurements were the ones reported on by Stenbaek-Nielsen et al. (1973) and Frank and Sigwarth (2003). Other conjugate images of the aurora (e.g. Motoba et al. 2010) did not come from calibrated cameras.

The effect of precipitating protons on auroral emissions and ionospheric ionization also depends on the inclination of the magnetic field lines (e.g. Synnes et al. 1998; Gérard et al. 2001). Energetic precipitating protons charge exchange with the ambient atmosphere repeatedly as they descend. Due to the large gyro radius compared to electrons, and the decoupling from the magnetic field as they pick up electrons to become neutral hydrogen, the energy of the protons is deposited over a much larger area than that threaded by their original field lines. The angle of incidence, the inclination angle of the magnetic field, partly determines where this energy is deposited (Fang et al. 2005). Thus ionization and heating from proton precipitation will vary with field inclination. Doppler-shifted emissions from hydrogen, which can be uniquely attributed to proton precipitation (Vegard 1939), also have some dependence on the inclination, since the Doppler shift depends on the line of sight relative to the path of the hydrogen. Since the path of the hydrogen is predominantly along magnetic field lines, the spectrum from such emissions can also be expected to depend on the inclination (Gérard et al. 2001). Although the inclination effect has been studied extensively by modelers, we are not aware of any observational study showing a longitudinal or hemispheric variation in the proton aurora which can be related to this effect. This may well be because of the rather modest differences in inclination, as seen in Fig. 2.

## 9 Concluding Remarks

In Sect. 2 we quantified the differences in the Earth's magnetic field between conjugate points in the ionosphere, and also the differences in sunlight exposure in the two magnetic hemispheres. We have shown that these differences, which can be significant, lead to asymmetries in ionospheric convection, thermospheric winds, currents and magnetic field perturbations, ion outflow, electron density, and auroral emissions. Several of these differences are not yet fully understood and should be a topic of research for years to come. As is clear from the extensive list of references, considerable work has been devoted to topics for which the field asymmetries are relevant, but only a minority directly address the asymmetry effects.

Differences in field strength and solar irradiance at conjugate hemispheres lead to different ionospheric manifestations of magnetospheric disturbances. Observations in both hemispheres give two views of the same magnetospheric disturbance, propagated to the ionosphere under different conditions. Analysis of hemispheric differences can therefore potentially elucidate the mechanisms involved in the magnetosphere–ionosphere–thermosphere coupling. The hemispheric differences thus represent an opportunity to study aspects of the magnetosphere–ionosphere coupling which would not be possible if the field was symmetrical. Fully exploiting this opportunity, and understanding the hemispheric differences reviewed in this paper, requires good data coverage from both hemispheres, as well as new approaches to analyze existing data, for example by novel data fusion techniques, and analyses which accurately take into account the differences in main field geometry in the two hemispheres.